\documentstyle[aps,preprint,tighten]{revtex}
\newcommand{\lsim}{\mathrel{\mathop{\kern 0pt \rlap
  {\raise.2ex\hbox{$<$}}}
  \lower.9ex\hbox{\kern-.190em $\sim$}}}
\newcommand{\gsim}{\mathrel{\mathop{\kern 0pt \rlap
  {\raise.2ex\hbox{$>$}}}
  \lower.9ex\hbox{\kern-.190em $\sim$}}}
\newcommand{\di}	{\mbox{d}}
\newcommand{\lm}	{\lambda}
\newcommand{\lrto}	{\leftrightarrow}
\newcommand{\no}	{\nonumber}
\newcommand{\om}	{\omega}
\newcommand{\vphi}	{\varphi}

\begin{document}
\draft
\preprint{
\begin{tabular}{r}
JHU-TIPAC-97020 
\\
KIAS-P97006
\\
SNUTP 97-138
\\
hep-ph/9710207
\end{tabular}
}
\title{Pulsar Velocity with Three-Neutrino Oscillations
\\
in Non-adiabatic Processes}
\author{
C.W. Kim$^{\mathrm{a,b}}$\footnote{E-mail: kim@eta.pha.jhu.edu}, 
J.D. Kim$^{\mathrm{a}}$ 
\footnote{E-mail: jdkim@eta.pha.jhu.edu} and
J. Song$^{\mathrm{c}}$\footnote{E-mail: jhsong@ctp.snu.ac.kr}
}
\address{
\begin{tabular}{c}
$^{\mathrm{a}}$Department of Physics and Astronomy, 
The Johns Hopkins University,
\\
Baltimore, Maryland 21218, USA
\\
$^{\mathrm{b}}$School of Physics, Korean Institute for Advanced Study,
Seoul 130-012, Korea
\\
$^{\mathrm{c}}$Center for Theoretical Physics,
Seoul National University, Seoul 151-742, Korea
\end{tabular}
}

\date{October 1,1997}
\maketitle
\begin{abstract}
We have studied the position dependence of neutrino energy
on the Kusenko-Segr\`{e} mechanism
as an explanation of the proper motion of pulsars.
The mechanism is also examined in three-generation mixing
of neutrinos and in a non-adiabatic case.
The position dependence of neutrino energy requires the 
higher value of magnetic field such as
$B\sim 3\times 10^{15}$ Gauss in order to explain
the observed proper motion of pulsars.
It is shown that possible non-adiabatic processes decrease 
the neutrino momentum asymmetry, whereas an excess of electron neutrino
flux over other flavor neutrino fluxes increases 
the neutrino momentum asymmetry.
It is also shown that a general treatment with all three
neutrinos does not modify the result of the two
generation treatment if the standard neutrino mass hierarchy
is assumed. 
\end{abstract}
\pacs{\em 14.60.Pq,\,97.60.Bw,\,97.60.Gb}

In a recent paper, Kusenko and Segr\`{e}~\cite{KS} have proposed a
new explanation for the peculiar velocities 
(or kick velocities) of pulsars
by considering magnetically distorted neutrino oscillations.
Many mechanisms\cite{Gott,Harrison,C,B} for the phenomena
have been suggested by some asymmetry during
the supernova collapsing and explosion.
However, most of them have difficulties
in explaining the observed average pulsar velocity $450\pm90$ km/s \cite{Lyne}.
On the other hand, the mechanism suggested in Ref.~\cite{KS},
hereafter called the KS mechanism, 
is successful~\cite{Birkel} for a reasonable value
of the magnetic field in pulsars, i.e., 
$B\sim10^{14\sim 15}$ Gauss.
The KS mechanism assumes that neutrino oscillations, 
$\nu_e\leftrightarrow\nu_{\tau}$, take
place between the electron- and tau-neutrinospheres,
where the matter density is $\rho\approx 10^{11\sim 12}$ g/cm$^3$.
Since neutrinos emitted during the cooling of
a protoneutron star have roughly 100 times the momentum
of the proper motion of the pulsar,
the neutrino momentum asymmetry of $\Delta k/k\sim 1$ \% can
explain the observed motion of pulsars.
Reference \cite{KS} shows that the magnetic field of
$3\times 10^{14}$ Gauss can produce the desirable momentum asymmetry. 
In his comment, Qian~\cite{Qian} pointed out that 
the approximate relation used in Ref.~\cite{KS}
is not a valid assumption
and that somewhat higher magnetic field is required in order to
explain the observed kick velocity.
The differences between Ref.~\cite{KS} and Ref.~\cite{Qian}
are in the flux factor $1/3$ and the estimate of the matter
density ($\rho= 10^{11}$ and $10^{12}$ g/cm$^3$)
besides the different expressions
for the ratio of scale heights $h_{N_e}/h_T$,
where $h_{N_e}\equiv|\di\ln N_e/\di r|^{-1}$ and
$h_T\equiv|\di\ln T/\di r|^{-1}$.

Therefore, it is worthwhile to re-examine the KS mechanism
in a more general case for supernovae and neutrinos.
In this paper, first we will consider the position dependence 
of neutrino energy inside the protoneutron star, which was
ignored in both Ref.~\cite{KS} and \cite{Qian}.
Our expression for the neutrino momentum asymmetry will be shown to 
reproduce the results in Ref.~\cite{Qian} and Ref.~\cite{KS},
respectively, under different approximations.
We will also study the treatment of 
the KS mechanism in a three-neutrino generation
scheme and see how the results are modified for possible
non-adiabatic processes. The possibility of different fluxes for
each neutrino flavor and its effect on the pulsar velocities are
also discussed.

First, let us briefly review the KS mechanism in which 
only $\nu_e\leftrightarrow\nu_{\tau}$ conversions are considered.
The difference in the mean free paths of $\nu_e$
and $\nu_{\tau}$ renders the $\nu_e$ neutrinosphere, $S_e$, to be
located at larger radius than the $\nu_{\tau}$ neutrinosphere, $S_{\tau}$.
If a resonant conversion between $\nu_e$ and $\nu_{\tau}$ 
occurs outside $S_{\tau}$ but inside $S_e$,
the $\nu_e$ converted from $\nu_{\tau}$ is 
absorbed or thermalized by the background
medium whereas the $\nu_{\tau}$ converted from $\nu_e$ 
escapes freely.
In effect, the $\nu_{\tau}$'s are expected to propagate out
not from the $S_{\tau}$, but from the resonance surface
which is distorted in the presence of magnetic field.
The resonance condition for two neutrino oscillations
in $\vec{B}$ field is \cite{ResonanceB}
\begin{eqnarray}
\label{eq:1}
\frac{\Delta_0 ^2}{2k(r)} \cos 2 \theta &=&
\sqrt{2} G_F N_e(r) 
+ 
\frac{e G_F}{\sqrt{2}} 
\left(
\frac{3 N_e(r)}{\pi^4}
	  \right)^{1 \over 3}
	\hat{k} \cdot \vec{B}
\\ \nonumber
&=&
\sqrt{2} G_F N_e 
\left[
1+
\frac{e }{{2}} \left(
   \frac{3 }{\pi^4 N_e^2}
				  \right)^{1 \over 3}
				B \cos\phi
\right]
\\ \no
&\equiv&
\sqrt{2} G_F N_e 
[1+\varepsilon_B],
\end{eqnarray}
where $\Delta_0 ^2 \equiv m(\nu_3)^2-m(\nu_1)^2$
in the case of $\nu_e\leftrightarrow\nu_{\tau}$ conversion,
$N_e=(n_{e^-}-n_{e^+})$,
$\theta$ is the neutrino mixing angle in vacuum,
$G_F$ the Fermi constant,
$\vec{B}$ the magnetic field of pulsars,
$k \simeq | \vec{k}|$ the energy of relativistic neutrinos,
$\phi$ the angle between $\vec{B}$ and $\vec{k}$,
and $r$ the radial coordinate.
Around typical neutrinospheres with $\rho \approx 10^{11\sim 12}$ g/cm$^3$
and the electron fraction per baryon $Y_e \approx 0.1$, 
the order of magnitude of $\varepsilon_B$
is estimated to be
\begin{equation}
\label{eq:2}
\varepsilon_B
\approx 
0.002 
\left(
\frac{Y_e}{0.1}
\right)^{-\frac{2}{3}}
\left(
\frac{\rho}{10^{12}\mbox{g/cm}^3}
\right)^{-\frac{2}{3}}
\left(
\frac{B}{10^{14}\mbox{Gauss}}
\right),
\end{equation}
which is indeed small so that a perturbation calculation
can be applied.
We caution the reader that the neutrino energy $k$ as well as
the electron number density $N_e$ depend on $r$.
The deeper the neutrinos are emitted from,
the higher their energies are.
To the zeroth order,
i.e. in the electron background without magnetic field,
the resonance condition is satisfied at $r_0$ as 
\begin{equation}
\label{eq:3}
\Delta_0 ^2 \cos 2 \theta = 2 \sqrt{2} G_F N_e(r_0) k(r_0).
\end{equation}
The presence of $\vec{B}$ field modifies the resonance position into
\begin{equation}
\label{eq:4}
r(\phi) =r_0+ \delta_r \cos \phi,
\end{equation}
where $\delta_r \ll r_0$ 
because $\varepsilon_B$, the correction due to the presence of
magnetic field, is
very small(see Eq.(\ref{eq:2})).
Substituting Eq.(\ref{eq:4}) into Eq.(\ref{eq:1})
and taking into account the $r$-dependence,
we obtain,
to the leading order in $\varepsilon_B$ or $\delta_r/r_0$,
\begin{equation}
\label{eq:5}
\left(
\left| \frac{\di \ln N_e}{\di r} \right|_{r_0} 
+
\left| \frac{\di \ln k}{\di r} \right|_{r_0} 
\right)
\delta_r = \frac{e}{2} 
\left(
\frac{3}{\pi^4 N_e^2}
\right)^{1\over 3}
B,
\end{equation}
where we have used the fact that 
$\di N_e/\di r <0$
and
$\di k/\di r <0$.
Under the assumption of blackbody radiation for neutrinos,
the energy flux of neutrinos ${\cal{F}}_\nu(r)$ is proportional to
$T^4$ so that 
$
|\di \ln k/\di r|
=4 
|\di \ln T/\di r|
$.
Defining the scale heights at the resonance point as
$h_T \equiv 
|\di \ln T/\di r|^{-1}_{r_0}$
and
$h_{N_e} \equiv 
|\di \ln {N_e}/\di r|^{-1}_{r_0}$
and using $N_e \approx \mu_e^3/3\pi^2$ where
$\mu_e$ is the chemical potential of the
degenerate relativistic background electrons,
Eq.(\ref{eq:5}) reduces to
\begin{equation}
\label{eq:6}
\delta_r =\frac{3eB}{2\mu_e^2} 
\frac{h_T h_{N_e}}{4 h_{N_e}+h_T}.
\end{equation}
The asymmetry of the total neutrino momentum distribution  is,
then, given by
\begin{eqnarray}
\label{eq:7}
\frac{\Delta k}{k} 
&=&
\frac{1}{6}
\frac{ 
\int_0^\pi {\cal{F}}_\nu(r_0+\delta_r \cos\phi)\cos\phi\sin\phi\di\phi
}{
\int_0^\pi {\cal{F}}_\nu(r_0+\delta_r \cos\phi)\sin\phi\di\phi
}
\\   \nonumber
 &\approx&
0.01 
\left(\frac{1}{1+(h_T/4h_{N_e})}\right)
\left(
\frac{B}{3 \times 10^{15}\mbox{Gauss}}
\right)
\left(
\frac{\mu_e}{11 \mbox{MeV}}
\right)^{-2}
,
\end{eqnarray}
where all six species of neutrinos are approximated to equally contribute
to the total energy of a supernova.
It is to be emphasized here that unity in the denominator
of the first bracket in Eq.(\ref{eq:7}) is due to 
the $r$-dependence of neutrino energy.
Note that we have not used the relation 
$\di N_e/\di T\approx (\partial N_e/\partial T)_{\mu_e}$
in Ref.\cite{KS}
which is, strictly speaking, not valid inside protoneutron stars 
as pointed out in Ref.\cite{Qian}.
The neglect of $r$-dependent neutrino energy,
which amounts to replacing $1/(1+(h_T/4h_{N_e}))$
by $4h_{N_e}/h_T$,
leads to
\begin{equation}
\label{eq:8}
\frac{\Delta k}{k} \approx 
  \frac{eB}{3\mu_e}\frac{h_{N_e}}{h_T}
= \frac{2}{9}\left(\frac{3}{2}
  \frac{eB}{\mu_e^2}\right)\frac{h_{N_e}}{h_T},
\end{equation}
which is the result in Ref.\cite{Qian}.
Further, if we ignore the temperature dependence of chemical
potential which justifies 
$\di N_e/\di T\approx (\partial N_e/\partial T)_{\mu_e}$, 
the ratio of the scale heights can be written as
$h_{N_e}/h_T\sim 3 N_e/2T^2\mu_e$. Then, using the relation
$N_e\sim \mu^3_e/3\pi^2$, we can rewrite Eq.(\ref{eq:8}) as
\begin{equation}
\label{eq:9}
\frac{\Delta k}{k} \approx \frac{1}{3}\left(\frac{e}{2\pi^2}
  \frac{B}{T^2}\right)
,
\end{equation}
which is the result in Ref.\cite{KS}
(with the correct flux factor $1/3$).
If we set $B\sim 10^{15}\mbox{Gauss}$,
$\rho\sim 10^{11}\mbox{g/cm}^3$,
$Y_e\sim 0.1 (\mu_e\sim 11\mbox{MeV})$ and 
$T\sim 3$MeV, we have $h_{N_e}/h_T\sim 0.7$ and therefore
$\Delta k/k\sim 0.01$.
However, numerical supernova calculations~\cite{Burrows}
show that the scale heights are $h_{N_e}\sim 6$ km,
$h_{\rho}\sim 7$ km, and $h_T\sim 22$ km at the stage
of the protoneutron star in a model of fixed baryon mass,
$1.4 M_{\odot}$, without accretion or convection.
Therefore, the ratio of the scale heights of the model\cite{Burrows}
is $h_{N_e}/h_T\sim 0.3$ which is slightly
different from the result, 0.7, of the approximate
relation $h_{N_e}/h_T\sim 3N_e/2T^2\mu_e$,
increasing the value of $B$ to have
$\Delta k/k\sim 0.01$.
This was the point stressed in Ref.~\cite{Qian}. 
If $h_{N_e}/h_T$ turns out to be large ($h_{N_e}/h_T>1$), then
the $r$-dependence of neutrino energy 
becomes relatively important and one has to resort to
a general formula in Eq.(\ref{eq:7}). In general, the value of 
$h_{N_e}/h_T$ depends on the model of the birth of
neutron star and on the time evolution of the collapse-driven
supernova explosion. 
In summary, the incorporation of the $r$-dependence of the
neutrino energy as manifested 
in the first parentheses in Eq.(\ref{eq:7})
requires, in general, higher $B$ values in order to explain the 
observed birth velocity of pulsars. 

Now, let us discuss the problem in a three-neutrino generation scheme
with a mass hierarchy that can accommodate the solar neutrino
data\cite{Bilenky95}.
According to a general theory of three
neutrino oscillations,
the weak eigenstates of neutrinos $\nu_{e,\mu,\tau}$
are superpositions of the mass eigenstates $\nu_{1,2,3}$
via a $3 \times 3$ unitary matrix $U$\cite{neutrino}:
\begin{equation}
\label{eq:10}
\left(
\begin{array}{c}
\nu_e \\ \nu_\mu \\ \nu_\tau
\end{array}
\right)
= U
\left(
\begin{array}{c}
\nu_1 \\ \nu_2 \\ \nu_3
\end{array}
\right).
\end{equation}
Assuming, for simplicity, that
$CP$ is conserved,
we adopt a useful parameterization of $U$ as
\begin{eqnarray}
\label{eq:11}
U &=& 
e^{i\psi \lm_7}
e^{i\vphi \lm_5}
e^{i\om \lm_2}
\\ \no
&=&
\left(
\begin{array}{ccc}
1 & 0 & 0 \\
0 & C_\psi & S_\psi \\
0 & -S_\psi & C_\psi
\end{array}
\right)
\left(
\begin{array}{ccc}
C_\vphi & 0 & S_\vphi \\
0 & 1 & 0 \\
-S_\vphi & 0 & C_\vphi
\end{array}
\right)
\left(
\begin{array}{ccc}
C_\om & S_\om & 0 \\
-S_\om &  C_\om & 0 \\
0 & 0 & 1 \\
\end{array}
\right)
,
\end{eqnarray}
where $C_\om=\cos \om$,
$S_\om=\sin\om$ and the $\lm$'s are the Gell-Mann matrices.
In an electron-rich medium, 
the equations of motion for the weak eigenstates of neutrinos are
\begin{equation}
\label{eq:12}
i \frac{\di }{\di t}
\left(
\begin{array}{c}
\nu_e \\ \nu_\mu \\ \nu_\tau
\end{array}
\right)
= \frac{1}{2E}\left[
U
\left(
\begin{array}{ccc}
m_1^2 
& 0 & 0 \\
0 &
m_2^2 
& 0 \\ 
0 & 0 &
m_3^2
\end{array}
\right)
U^\dagger +
\left(
\begin{array}{ccc}
A & 0 & 0 \\
0 & 0 & 0 \\ 
0 & 0 & 0
\end{array}
\right)
\right]
\left(
\begin{array}{c}
\nu_e \\ \nu_\mu \\ \nu_\tau
\end{array}
\right)
,
\end{equation}
where $A$ is the matter-induced mass squared of $\nu_e$
which is, in the presence of magnetic field,
\begin{equation}
\label{eq:13}
A = 2\sqrt{2} G_F N_e E + \sqrt{2}e G_F 
\left(
\frac{3 N_e(r)}{\pi^4}
	  \right)^{1 \over 3}
	\vec{k} \cdot \vec{B}
.
\end{equation}

We will consider the pulsar kick velocity by three 
neutrino mixing under two natural conditions:
(1) $m_1 \ll m_2 \ll m_3$;
(2) the mixing angles in vacuum are small.
It is easy to see then that three neutrino oscillations in matter
separate into two parts,
each being similar to two neutrino oscillations in matter.
Two resonant conversions can take place at well separated locations.
We call $R_h$ the resonance surface which occurs at higher density,
and $R_l$ the one at lower density.
The conditions for each resonance are\cite{Kuo}
\begin{eqnarray}
\label{eq:14}
A_{R_h}  &\cong&
\left[
m_3^2-\frac{1}{2}
\left\{
m_1^2 (1+C_\om)
+
m_2^2 (1-C_\om)
\right\}
\right]
C_{2\vphi}
,
\\ \no
A_{R_l} &\cong&
(m_2^2-m_1^2) C_{2\om}
.
\end{eqnarray}
In a three generation scheme, therefore,
the relative positions of two resonance surfaces to
neutrinospheres are essential in order to study the effects of
neutrino oscillations on the birth velocities of
pulsars.
First, let us discuss the relative positions of three neutrinospheres.
A neutrinosphere may be considered as an imaginary surface such that
inside the surface 
neutrinos are trapped due to the interactions with medium,
but outside they escape freely.  Therefore, the larger the scattering
cross section is, the outer the neutrinosphere is located.
Since the temperature ($T\approx 3$ MeV) around the neutrinospheres
is too low to thermally create $\mu$ or $\tau$,
the $\nu_{\mu}$ neutrinosphere is expected to coincide with the
$\nu_{\tau}$ neutrinosphere, $S_{\tau}$.
As can be seen in Eq.(\ref{eq:14}), the locations of resonance surfaces are
dependent on the masses and the mixing angles of neutrinos in vacuum which
have been restricted by various measurements and observations.
According to a comprehensive analysis of solar and
atmospheric neutrino experiments
in a hierarchical three-generation scheme, 
$m_1 \ll m_2 \ll m_3$,
the Mikheyev-Smirnov-Wolfenstein (MSW) solution to the 
solar neutrino deficit problem suggests 
$m_2^2-m_1^2\approx 10^{-5}$eV$^2$~\cite{solar} and
the atmospheric neutrino anomaly leads to 
$m_3^2 \gsim 10^{-3}$eV$^2$~\cite{Fogli}.
Therefore,
the density around neutrinospheres with 
$\rho\approx 10^{11\sim 12}$ g/cm$^3$ is too high for the lower
resonance to occur, implying that
the lower resonance surface $R_l$ lies outside the $S_e$.
Within the current experimental limit of $m_3$,
the $R_h$ (the resonant conversion 
$\nu_e \lrto \nu_\tau$\footnote{
Of course, this is an approximate interpretation,
which is valid when the neutrino mixing angles
in vacuum are very small.})
can lie between the $S_e$ and the $S_{\tau}$
\cite{ParticleData}.

In the following we assume
that the $R_l$ for $\nu_e \lrto \nu_\mu$
is located outside the $S_e$, but the $R_h$ for
$\nu_e \lrto \nu_\tau$ between the $S_e$ and the $S_{\tau}$.
At the $R_h$, $\nu_\tau$ produced inside the $S_{\tau}$ will
be converted into $\nu_e$ and absorbed,
whereas $\nu_e$ produced inside the $S_{\tau}$ will turn into
$\nu_\tau$ and will propagate out with the collapsing energy
of the protoneutron star.
The $\nu_\mu$ does not undergo any resonant conversion at the $R_h$.
As neutrinos pass the lower resonance surface,
neutrino flavors are resonantly converted again
on the magnetically distorted surface.
However,
this resonant conversion outside the $S_e$
does not contribute to the pulsar velocities
because the momentum transfer of neutrinos is independent of
whether or not the neutrino flavor is changed.
This is in contrast to the case where the resonance surface lies 
between the $S_e$ and the $S_{\tau}$ so that $\nu_{\tau}$
converted from $\nu_e$ propagates out 
but $\nu_e$ from $\nu_{\tau}$ cannot,
leading to an asymmetry in the neutrino momentum distribution.
Once neutrinos  escape from the $S_e$,
therefore,
flavor-changing of the neutrinos does not
affect the kick velocity of pulsars.
We summarize that if the value of $(m_2^2-m_1^2)$ is
restricted by the MSW solution to the solar neutrino problem
so that the lower resonance occurs outside the $S_e$,
the consideration of three neutrino oscillations
does not affect the result of 
the KS mechanism based on the two generation scheme.

Next, we discuss the effects of possible
non-adiabatic neutrino oscillations
on the birth velocity of pulsars.
It is, of course, obvious that 
non-adiabatic processes make
$\nu_e\rightarrow\nu_{\tau}$
conversion less efficient, reducing the kick velocity.
We wish to study this in some detail.
Let ${\cal P}(\nu_\alpha \rightarrow \nu_\beta)$ 
be a general probability for $\nu_\alpha$ produced deep 
inside the resonance surface 
to be observed as $\nu_\beta$ outside the resonance surface.
In adiabatic cases where instantaneous effective mass
eigenstates approximate the energy eigenstates of Hamiltonian,
the transition probability 
${\cal P}_{\rm A}(\nu_\alpha \rightarrow \nu_\beta)$~\cite{Kuo88} is
\begin{equation}
\label{eq:15}
{\cal P}_{\rm A}(\nu_\alpha \rightarrow \nu_\beta)
= \langle \nu_\beta |
\left(
\begin{array}{cc}
C_f^2 & S_f^2 \\ S_f^2 & C_f^2
\end{array}
\right)
\left(
\begin{array}{cc}
1 & 0 \\ 0 & 1
\end{array}
\right)
\left(
\begin{array}{cc}
C_i^2 & S_i^2 \\ S_i^2 & C_i^2
\end{array}
\right)
|\nu_\alpha\rangle,
\end{equation}
where $C_{f,i}=\cos \theta_{f,i}$ and $\theta_f(\theta_i)$
is the effective mixing angle of neutrinos in matter
after (before) neutrinos go through the resonance surface
and
\begin{equation}
\label{eq:16}
|\nu_e\rangle
=
\left(
\begin{array}{c}
1 \\ 0
\end{array}
\right)
, ~~
|\nu_\tau\rangle
=
\left(
\begin{array}{c}
0 \\ 1
\end{array}
\right)
.
\end{equation}
If the oscillation process is non-adiabatic,
a level-crossing 
between the effective mass eigenstates takes place,
particularly near the resonance point\cite{neutrino}.
With the level-crossing (Landau-Zener) probability $P_{LZ}$, 
we have the following non-adiabatic transition probabilities
\begin{equation}
\label{eq:17}
{\cal P}_{\rm NA}(\nu_\alpha \rightarrow \nu_\beta)
= \langle \nu_\beta |
\left(
\begin{array}{cc}
C_f^2 & S_f^2 \\ S_f^2 & C_f^2
\end{array}
\right)
\left(
\begin{array}{cc}
1-P_{LZ} & P_{LZ} \\ P_{LZ} & 1-P_{LZ}
\end{array}
\right)
\left(
\begin{array}{cc}
C_i^2 & S_i^2 \\ S_i^2 & C_i^2
\end{array}
\right)
|\nu_\alpha\rangle.
\end{equation}
The level-crossing probability is $P_{LZ}=\exp(-\frac{\pi}{4}Q)$
where the adiabaticity parameter $Q$ is given by
$Q=\Delta_0^2S^2_{2\theta}
h_{N_e}/kC_{2\theta}$ in the two generation case. 
For $\Delta_0^2\sim 10^4$ eV$^2$, $h_{N_e}\sim 6$ km,
and $k\sim 10$ MeV, the non-adiabatic effect cannot be
ignored when the vacuum mixing angle $\theta$ is smaller
than $10^{-4}$ (Recall that the mixing angle of relevance
is $\nu_e-\nu_{\tau}$ mixing angle, therefore, a very 
small mixing angle is not ruled out). 
For example, the level-crossing probability is
$P_{LZ}\sim 0.38$ for $\theta=10^{-4}$.
In the extremely non-adiabatic case,  
flavor changing processes would not take place, 
so that the KS mechanism becomes inoperative. 

Let us consider a general non-adiabatic case
with a possibility that the fluxes of $\nu_e$'s and $\nu_{\tau}$'s
from the protoneutron star are different.
The collapsing supernova core has about $10^{57}$ protons
which produce neutrinos via $p+e^{-}\rightarrow n+\nu_{e}$, 
resulting in the formation of a neutron star.
Since each $\nu_e$ emitted from the core 
 carries away an average energy 10 MeV, about $10^{52}$ ergs
are emitted by $\nu_e$'s during the neutronization processes.
This is less than 5\% of all the $\nu$'s radiated\cite{Mayle}.
The rest of the neutrinos come from pair
processes such as 
$e^{+}+e^{-}\rightarrow\nu_i+\bar{\nu}_i$,
where $i=e,\mu$ or $\tau$. 
Since $\nu_{\mu}$ and $\nu_{\tau}$ are produced via neutral
currents but $\nu_e$ via both charged and neutral currents, 
there exist some differences between the $\nu_e$ flux
and other neutrino fluxes, defined collectively by $\nu_{\beta}$. 
We will introduce the parameter $\alpha_{\nu_e}$ 
such that neutrino fluxes are given by 
${\cal F}_{\nu_e}=(1+\alpha_{\nu_e}){\cal F}$
and ${\cal F}_{\nu_{\beta}}={\cal F}$, where ${\cal F}$ is 
the common flux factor. 
The existence of another enhancing channel for  $\nu_e$ production
implies $\alpha_{\nu_e}>0$.
The asymmetry in the neutrino momentum distribution due to
non-adiabatic oscillations is then given by 
\begin{eqnarray}
\label{eq:18}
\left( \frac{\Delta k}{k} \right)_{\rm NA} 
&\approx &
\frac{
{\cal P}_{\rm NA}(\nu_e\rightarrow\nu_{\tau})(1+\alpha_{\nu_e})
}{
(6+\alpha_{\nu_e})
}
\frac{\int_0^{\pi}{\cal F}(r_0+\delta_r\cos\phi)
\cos\phi\sin\phi\di\phi}{\int_0^{\pi}{\cal F}(r_0+\delta_r\cos\phi)
\sin\phi\di\phi}   \\ \nonumber
&\approx &
\left( \frac{\Delta k}{k} \right) 
{\cal P}_{\rm NA} (\nu_e\rightarrow\nu_{\tau})
\left(1+\frac{5}{6}\alpha_{\nu_e}\right)
,
\end{eqnarray}
where the second approximation is valid
for small value of $\alpha_{\nu_e}$.
The term $(\Delta k/k)$ on the right hand side of Eq.(\ref{eq:18})
is given by Eq.(\ref{eq:7}).
In the small vacuum mixing case ($\theta_f\sim 0,\theta_i\sim \pi/2$)
the transition probability ${\cal P}_{\rm NA}(\nu_e\rightarrow\nu_{\tau})$
can be further approximated, resulting in
\begin{equation}
\label{eq:19}
\left(\frac{\Delta k}{k} \right)_{\rm NA} \approx
\left(\frac{\Delta k}{k} \right) (1-P_{LZ})
\left(1+\frac{5}{6}\alpha_{\nu_e}\right).
\end{equation}
This shows that the non-adiabaticity has a tendency to
decrease the asymmetry in the neutrino momentum distribution.
Equation (\ref{eq:19}) shows that 
extremely non-adiabatic processes ($P_{LZ}=1$)
cause  no kick velocity due to the absence of flavor conversion,
as discussed before.
The pulsar velocity increases with $\alpha_{\nu_e}$.
This is also expected because the more $\nu_e$'s are 
produced inside the core, the more they turn to the $\nu_\tau$'s,
leading to a larger birth velocity of pulsar.
Therefore, the effect of non-adiabaticity  is opposite to
that of $\alpha_{\nu_e}>0$.
When we set, for example, the vacuum mixing angle 
$\theta=1.6\times 10^{-4}$ and $\alpha = 10$ \%, 
the contributions of the non-adiabaticity and the excess 
flux of electron neutrinos to the momentum asymmetry are almost canceled out.
This cancelation depends very sensitively on the value of $\theta$.

In summary, we have studied the Kusenko-Segr\`{e} mechanism
in a more general condition than in Refs.~\cite{KS} and \cite{Qian}. 
First, the position dependence of neutrino energy 
has been taken into account
in the calculation of the neutrino momentum asymmetry 
to explain the peculiar
pulsar velocity. This aspect becomes relatively important
in the case $h_{N_e}/h_T>1$.
In addition, we have clarified the difference between the two
treatments in Refs.~\cite{KS} and \cite{Qian}.
The KS mechanism is also studied in a three generation scheme
and in a non-adiabatic case.
If the solar neutrino and atmospheric neutrino problems are 
to be solved with the standard mass hierarchy, a generalization
of the treatment in Ref.~\cite{KS} in a three-generation
scheme leads to no new modifications.
The non-adiabaticity has a general tendency to decrease the kick
velocity whereas a larger electron neutrino flux compared to that of
other flavor gives rise to a higher momentum asymmetry.

\vspace{1cm}
{\bf Acknowledgments.}
JDK acknowledges support from the Korean Science and
Engineering Foundation (KOSEF) and
JS acknowledges support from the Korean Science
and Engineering Foundation (KOSEF) 
through the Center for Theoretical Physics (CTP).

\end{document}